\documentclass{article}

\newcommand{\paperTitle}{UNet\#: A UNet-like Redesigning Skip Connections for Medical Image Segmentation}

\PassOptionsToPackage{round, sort}{natbib}
\usepackage[preprint]{neurips_2022}

\usepackage[utf8]{inputenc} 
\usepackage[T1]{fontenc}    
\usepackage{hyperref}       
\usepackage{url}            
\usepackage{booktabs}       
\usepackage{amsfonts}       
\usepackage{nicefrac}       
\usepackage{microtype}      
\usepackage[table,xcdraw]{xcolor}
\usepackage{xspace}
\usepackage{multirow}
\usepackage{amsmath,amsopn}
\usepackage{xstring}
\usepackage{breqn}
\usepackage[capitalize,noabbrev,nameinlink]{cleveref}
\usepackage{dsfont}
\usepackage{graphicx}
\usepackage{floatrow}
\usepackage{blindtext}
\usepackage[font=small,labelfont=bf,tableposition=top]{caption}
\usepackage{subcaption}
\usepackage{varwidth}
\usepackage{multirow}
\usepackage[normalem]{ulem}
\usepackage{booktabs}

\newcommand{\PP}[1]{
    \vspace{2px}
    \noindent{\bf \IfEndWith{#1}{.}{#1}{#1.}}
    }

\newcommand{\squishitemize}{
\begin{list}{$\bullet$}
	{ \setlength{\itemsep}{0pt}
		\setlength{\parsep}{3pt}
		\setlength{\topsep}{3pt}
		\setlength{\partopsep}{0pt}
		\setlength{\leftmargin}{1.95em}
		\setlength{\labelwidth}{1.5em}
		\setlength{\labelsep}{0.5em} } }

\newcounter{Lcount}
\newcommand{\squishlist}{
	\begin{list}{\arabic{Lcount}. }
		{ \usecounter{Lcount}
			\setlength{\itemsep}{0pt}
			\setlength{\parsep}{3pt}
			\setlength{\topsep}{3pt}
			\setlength{\partopsep}{0pt}
			\setlength{\leftmargin}{2em}
			\setlength{\labelwidth}{1.5em}
			\setlength{\labelsep}{0.5em} } }
	
\newcommand{\squishend}{\end{list}}

\newfloatcommand{capbtabbox}{table}[][\FBwidth]
\useunder{\uline}{\ul}{}

\title{\paperTitle}

\author{%
  Ledan Qian \\
  College of Mathematics and Physics\\
  Wenzhou University\\
  Wenzhou, China 325025 \\
  \texttt{00802005@wzu.edu.cn} \\
  \And
  Xiao Zhou \\
  Information Technology Center\\
  Wenzhou University\\
  Wenzhou, China 325025 \\
  \texttt{00802002@wzu.edu.cn} \\
   \And
  Yi Li \\
  College of Computer Science and Artificial Intelligence\\
  Wenzhou University\\
  Wenzhou, China 325025 \\
  \texttt{liyi@wzu.edu.cn} \\
   \And
  Zhongyi Hu \\
  College of Computer Science and Artificial Intelligence\\
  Wenzhou University\\
  Wenzhou, China 325025 \\
  \texttt{huzhongyi@wzu.edu.cn} \\
}

\begin{document}

\maketitle

\begin{abstract}
As an essential prerequisite for developing a medical intelligent assistant system, medical image segmentation has received extensive research and concentration from the neural network community. A series of UNet-like networks with encoder-decoder architecture has achieved extraordinary success, in which UNet2+ and UNet3+ redesign skip connections, respectively proposing dense skip connection and full-scale skip connection and dramatically improving compared with UNet in medical image segmentation. However, UNet2+ lacks sufficient information explored from the full scale, which will affect the learning of organs' location and boundary. Although UNet3+ can obtain the full-scale aggregation feature map, owing to the small number of neurons in the structure, it does not satisfy the segmentation of tiny objects when the number of samples is small. This paper proposes a novel network structure combining dense skip connections and full-scale skip connections, named UNet-sharp (UNet\#) for its shape similar to symbol \#. The proposed UNet\# can aggregate feature maps of different scales in the decoder sub-network and capture fine-grained details and coarse-grained semantics from the full scale, which benefits learning the exact location and accurately segmenting the boundary of organs or lesions. We perform deep supervision for model pruning to speed up testing and make it possible for the model to run on mobile devices; furthermore, designing two classification-guided modules to reduce false positives achieves more accurate segmentation results. Various experiments of semantic segmentation and instance segmentation on different modalities (EM, CT, MRI) and dimensions (2D, 3D) datasets, including the nuclei, brain tumor, liver and lung, demonstrate that the proposed method outperforms state-of-the-art models.  
\end{abstract}
  
\section{Introduction}
\label{sec:intro}
Computer-aided diagnosis(CAD) system plays a vital role in disease diagnosis and treatment \citep{1}. Medical image segmentation is the first step of all tasks for the medical image processing task based on a CAD system. Essentially, it classifies the input medical images with pixel granularity, aiming to make the human tissue or pathological structure more extinct and intuitive. Medical workers can model the relevant organizations by using the segmentation results for subsequent applications. Medical image segmentation is a crucial significance in providing noninvasive information about the human structure. It is also helpful for radiologists to visualize the anatomical structure, simulate biological processes, locate pathological tissues, track the progress of diseases, and provide the information needed to evaluate radiotherapy or surgery \citep{2}.

Medical images are the discrete image representation generated by sampling or reconstructing. Its values mapping to different spatial locations can reflect the anatomical structure or functional organization of the human body. Medical images include computed tomography (CT), Ultrasound (US), Magnetic Resonance Imaging (MRI), X-ray, etc. These images are different from natural images with obvious boundaries, showing human tissue or lesion area obtained from professional imaging instruments; accordingly, medical images have the unclear edge contour of organs and tissues and complex brightness changes. The application of segmentation technology in these two images consequently presents different performance requirements. It is worth mentioning that medical images have higher standards for segmentation details.

The emergence and development of deep learning technology have effectively solved many issues encountered in medical image segmentation. Using the powerful learning ability of the deep neural network can abstract lesions or organ targets in high-dimensional through nonlinear transformation and in-depth extract the high-level semantic information of medical images layer by layer. This information describes the internal attributes of lesions or organs and can more accurately segment the image region of interest (ROI) automatically in medical images. UNet \citep{11} is the most classic application case in medical image segmentation. Recently, it has been used as a benchmark network to investigate a series of novel strategies, such as the introduction of ResNet and DesNet structural blocks \citep{12,13}, the development of attention mechanism \citep{14,15}, the reconstruction of convolution mode  \citep{16}, and the redesign of skip connection \citep{17,18,19}.

\begin{figure*}[t]
	\centering
	\vspace{-0.4cm}    
	\includegraphics[width=1\linewidth]{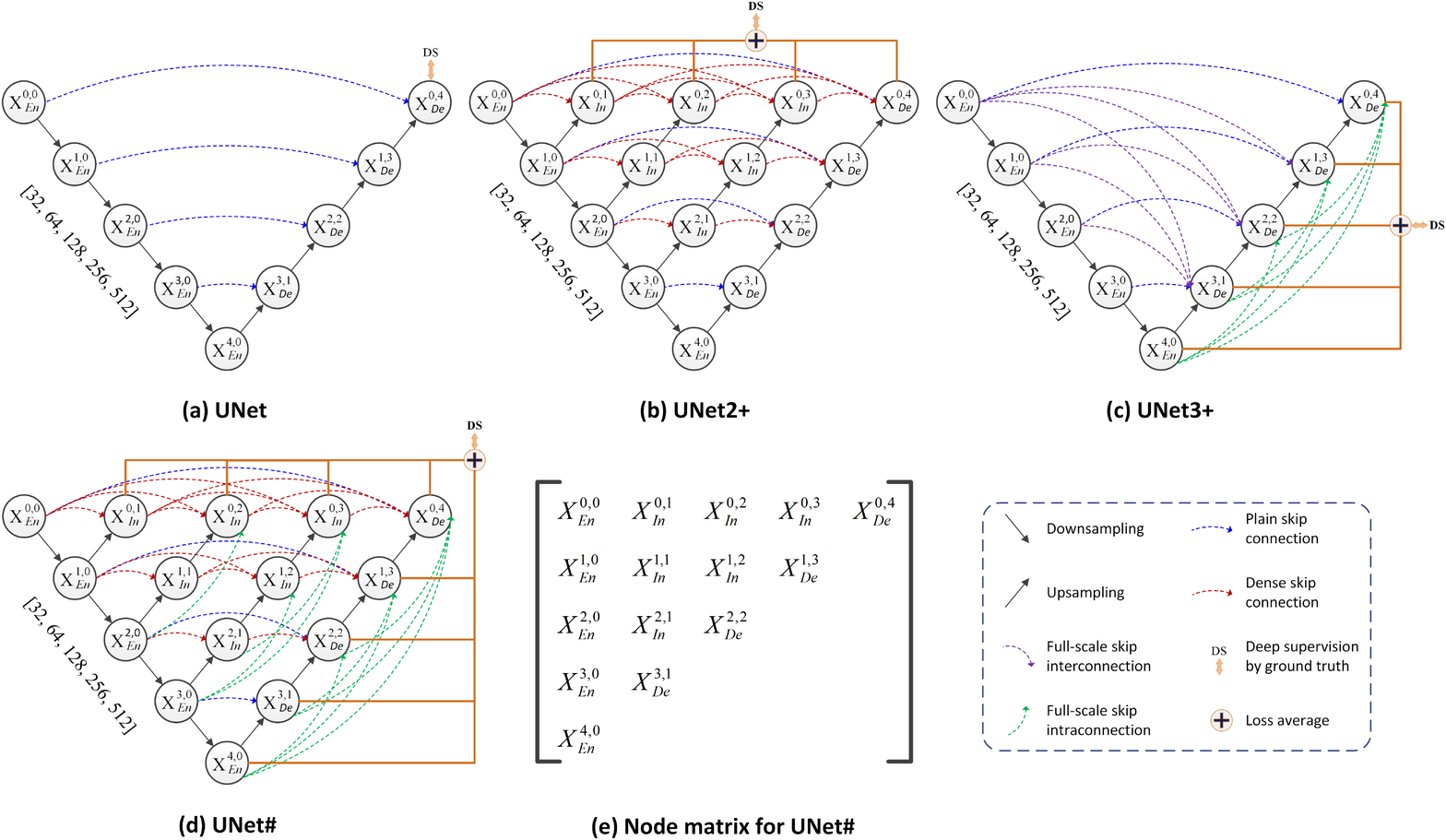}
	\caption{The model architecture comparison between UNet(a), UNet2+(b), UNet3+(c) and UNet\#(d). The elements of node matrix(e) composed of the UNet\#'s units shown as (d) are expressed by Eq. \ref{eq1}, easily comprehending the calculation method of the proposed UNet\#. The number of feature channels from each stage in backbone is 32, 64, 128, 256 and 512. The implementation of deep supervision in UNet\# is different from that in UNet2+ and UNet3+, which is detailed in Sec. \ref{sec4.2}.}
	\label{fig1}
\end{figure*}

Benefiting from the redesign of the skip connection, UNet2+ \citep{18} and UNet3+ \citep{19} have brought significant performance improvement compared with UNet. Nevertheless, it can not be ignored that UNet2+ lacks sufficient, comprehensive information exploring from the full scale, which makes it unable to learn the location and boundary of organs distinctly. Due to the fewer neurons in the UNet3+ structure, the segmentation of small targets is not ideal for the training with a small number of samples. Inspired by these two networks, we propose a novel structure of UNet\_like shown in Figure \ref{fig1} (d), which is very similar to the grid structure of \# symbols, named UNet-sharp (UNet\#). UNet\# has the advantages of UNet2+ and UNet3+, which can aggregate feature maps of different scales in the encoder sub-network and capture fine-grained details incorporating coarse-grained semantics from the full-scale feature map.

In addition, we add the loss function to the four decoder branch outputs, comparing with the post-downscale ground truth to supervise the model training, which further learn the hierarchical representation from the full-scale aggregation feature map. Moreover, introducing deep supervision to the other four middle layers and the final decoder branch with the same scale as the input image can realize the pruning function of the model. To suppress the false-positive problem further, we develop a classification-guided module (CGM) and apply it to all eight branch outputs to promote a more accurate segmentation for lesions and organs in medical images.

\PP{Contributions.}
Our contributions are as follows:
\squishlist
\item Proposing A novel medical image segmentation structure UNet\#. 
\item Adding deep supervision into the decoder branches' output to further capture coarse-grained and fine-grained semantic information from the full scale, and introducing deep supervision into the first layer nodes (including the final output node) can guide the model pruning.	
\item Designing two classification-guided modules for different output branches to achieve higher precision organ and lesion segmentation.
\item Carrying out extensive experiments on the different data sets to verify the consistent improvement of the method proposed in this paper from multiple aspects. 
\squishend
\section{Methods}
\label{sec4}
The redesigned nested and dense skip connections for UNet2+ make full use of the semantics from the feature maps at different layers. In UNet3+, the proposed full-scale skip connections can obtain enough information from the full-scale feature map. The network we designed combines the dense skip connection and full-scale skip connection to learn enough full-scale semantics while incorporating low-level details so that the model can yield more accurate segmentation results.

\begin{figure*}[tbp]
	\centering
	\includegraphics[width=1\linewidth]{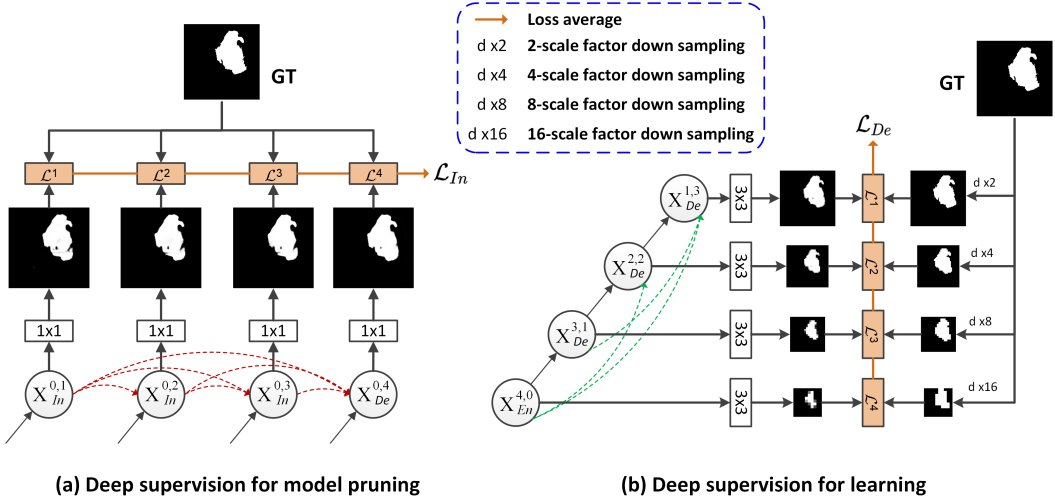}
	\caption{The schematic diagram of deep supervision in the proposed UNet\#. Deep supervision in our model has tow functions: (a) model pruning and (b) improving the learning skill of hierarchical representations. Adding loss function to the branches followed by $1 \times 1$ or $3 \times 3$ convolution operation produces the loss $\mathcal{L}^{i}, i \in(1\sim 4)$. $\mathcal{L}_{In}$ is used for supervision of model pruning, and $\mathcal{L}_{De}$ is used to supervise the improvement of the feature learning for the model. Most notably, the implementation methods of these two deep supervision (a) and (b) are not the same. The ground truth (GT) is from BraTs19 dataset (brain tumor). The \textit{MaxPool2d} operation is used for down sampling (d x2, d x4, d x8, d x16).}
	\label{fig2}
\end{figure*}

\subsection{Re-designed skip connections}
\label{sec4.1}
Skip connections \citep{71} play an influential role in improving the model's performance in the deep neural network. It can connect the shallow layer with the deep layer to retain the low-level features, avoiding the model performance degradation when adding multiple layers. We redesign the skip connections in the proposed network, interconnecting the features of each level encoder extracted through a dense revolution block, besides intraconnecting the deep and shallow features via full-scale skip connections. Furthermore, we also perform intraconnection through full-scale skip connections for decoders at different levels. Compared with the state-of-the-art network models (UNet, UNet2+, UNet3+), this redesigned skip connection method has a stronger full-scale information exploration ability and receptive fields of diverse sizes, which can realize high-precision segmentation organs with different sizes.

The proposed model structure is shown in Figure \ref{fig1} (d). As a UNet-like network, our model is similar to the UNet2+ design, adding six intermediate units, but the skip connections are redesigned. To better understand the model structure, a matrix, as shown in figure \ref{fig1} (e), composing 5-row and 5-columns cell nodes is used to abstractly represent the model structure units. The first column nodes of the matrix are the model's encoder, marked as $X_{En}^{i,0}, i \in [0,1,2,3,4]$. The calculation for the second column unit module is conducted as follows: firstly, upsamples the encoder feature of the deeper level with a two-scale factor; secondly, concatenates with the encoder of the same level on the channel, and finally outputs the final result of the unit using convolution operation. For instance, the encoder unit $X_{In}^{0,1}$ is obtained by the convolution calculation after operating the channel concatenation between the same level layer encoder $X_{En}^{0,0}$ and the 2-times upsampling of the deeper layer encoder $X_{En}^{1,0}$. The other column nodes of the matrix are calculated from two aspects: on the one hand, the feature information of the same level is thoroughly obtained through dense skip connections (interconnections). On the other hand, a chain of intra-decoder skip connections (intraconnections) transmits the high-level semantic information from the smaller-scale decoder layer. For example, the encoder $X_{In}^{0,2}$ is calculated via the convolution operation following the channel concatenation of the four feature maps consisting of dense skip interconnections from $X_{En}^{0,0}$, $X_{In}^{0,1}$, full-scale skip intraconnection from $X_{En}^{2,0}$, and the two-scale factor upsampling of $X_{In}^{1,1}$. The final output of the model, decoder $X_{De}^{0,4}$, is achieved by the convolution operation and the channel concatenation of the eight feature maps composed of dense skip interconnections from $X_{En}^{0,0}$, $X_{In}^{0,1}$, $X_{In}^{0,2}$, $X_{In}^{0,3}$ full-scale skip intraconnections from $X_{En}^{4,0}$, $X_{De}^{3,1}$, $X_{De}^{2,2}$ and the two-scale factor upsampling of $X_{De}^{1,3}$. The redesigned skip connections make the network more similar between encoder and decoder features at the semantic level than UNet2+ and UNet3+. This similarity can make the optimizer easier to optimize during computation, but also make the model more capable of full-scale feature information exploration from the full-scale aggregated feature maps.

The calculation of each model unit in the matrix is expressed as Eq. \ref{eq1}.

\begin{footnotesize} 
	\begin{equation}
		A^{I, J}=\left\{\begin{array}{ll}
			f^{2}\left(p\left(A^{I, 0}\right)\right), & J=0 \\
			f^{2}\left(\left[A^{I, 0}, u\left(A^{I+1,0}\right)\right]\right), & J=1 \\
			f^{2}\left(\left[\left[A^{I, j}\right]_{j=0}^{J-1}, u\left(A^{I+1, J-1}\right),\left[f\left(u^{J-j}\left(A^{i, j}\right)\right)\right]_{i=I+J, j=0}^{I+2, J-2}\right]\right), & J>1
		\end{array}\right.
		\label{eq1}
	\end{equation}
\end{footnotesize}

Where $A^{I,J}$ $ (I=[0,1,2,3,4]) $ represents the calculation results of each model unit. When $ J = 0 $, the first sub-formula defines the calculation method for the first column of node matrix, that is, the calculation method of the model encoder $ X_{En}^{i,0}, i \in [0,1,2,3,4] $. When $ J = 1 $, the second sub-formula is the calculation method of the second column of the matrix. When $ J> 1 $ $ (J\in[2,3,4]) $, the third sub-formula is the calculation method of the matrix's 3, 4, and 5 columns. $ P(\cdot) $ indicates $ 2\times2 $ \textit{Maxpool2d} operation meaning downsampling. $ u(\cdot) $ means to upsample on $ 2\times $, and $ u^{n}(\cdot)$ means $ 2^{n} $ times upsampling. $ [\cdot] $ denotes that the feature maps of each unit are concatenated on the channel; $ f(\cdot) $ represents 1-time sequential operation including \textit{Conv2d}, \textit{BatchNorm2d}, and \textit{ReLU} activation function, $ f^{n}(\cdot) $ indicates n-times sequential operations.

\subsection{Deep supervision}
\label{sec4.2}
As a training trick, deep supervision was proposed in DSN (Deeply-Supervised Nets) \citep{30} in 2014, aiming to get more totally training in shallow layers, which can avoid gradient disappearance and slow convergence. Compared with the conventional deep learning mechanism, deep supervision outputs the result \textit{out} at the terminal branch of the network while obtaining the result \textit{out\_m} with the same size after operating deconvolution or upsampling on the middle feature map of the network, sequentially, combining \textit{out\_m} and \textit{out} to train the network jointly. According to this idea, deep supervision applied in UNet2+ and UNet3+ achieve better results than the original structure.

We introduce deep supervision into the proposed model structure to make it operate in two modes \citep{17}: 1) accurate mode, wherein the branch output of the model will be averaged to calculate loss; 2) fast mode, which selects the branch with the best result, actually prunes the model (see Sec. \ref{sec4.4} for details) to improve the speed. As shown in Figure \ref{fig2}, we add the loss function to the eight branches' output of the proposed UNet\#. Specifically, implementation is as follows. The first layer feature maps ${X_{In}^{0,j}, j \in [1,2,3]}$ and $X_{De}^{0,4}$ generated by dense skip interconnections and full-scale skip intraconnections at multiple semantic levels, are followed by the operation of $ 1\times1 $ convolution and \textit{ReLU} activation function, producing the output results. Then comparing the results with GT calculates the loss $\mathcal{L}_{In}$ used to supervise the model for pruning. Additionally, the decoder $X_{De}^{1,3}$, $X_{De}^{2,2}$, $X_{De}^{3,1}$, $X_{En}^{4,0}$ branches followed by a $ 3\times3 $ convolution and \textit{ReLU} activation function bring the output results. Subsequently, respectively comparing with the GT of the corresponding size after down-scaling (consistent with the size of each output result) calculates the loss value $\mathcal{L}_{De}$, realizing the supervision of the model to the improvement of skills for learning hierarchical representations from full-scale feature maps.

Here, we use a mixed loss to calculate the loss value of each branch, which is composed of focal loss \citep{72}, Laplace smoothing dice loss, and Lovasz hinge loss \citep{73}. Mixing loss can bring smooth gradient and handling of class imbalance, which can capture both large-scale and delicate structures with clear boundaries. The mixed segmentation loss $\ell_{seg}$ is defined as follows:

\begin{equation}
	\ell_{seg}=0.5 * \ell_{f l}+\ell_{l s d}+0.5 * \ell_{l h}
	\label{eq2}
\end{equation}

Where $\ell_{f l}$ is focal loss, defined as follows:

\begin{equation}
	\ell_{f l}=-\alpha(1-\hat{Y})^{\beta} \log (\hat{Y})
	\label{eq3}
\end{equation}

Where $ \alpha $ and $ \beta $ are parameters, $ \alpha $ can control the shared weight of positive and negative samples to the total loss, $ \beta $ controls the weight of easy classify and difficult classify samples, and Y represents the predicted samples. In the experiment, we specify $ \alpha = 0.25 $ and $ \beta = 2 $ according to the setting with the best experimental effect in [4].

$\ell_{l s d}$ from Eq. \ref{eq2} is the soft Dice coefficient loss after Laplace smoothing, which can avoid the problem of division by 0 and overfitting, defined as follows:

\begin{equation}
	\ell_{l s d}=1-\frac{2 \cdot Y \cdot \hat{Y}+1}{Y+\hat{Y}+1}
	\label{eq4}
\end{equation}

Where $ Y $ is the label, $ \hat{Y} $ is the predicted sample.

$ \ell_{l h} $ from Eq. \ref{eq2} is Lovasz hinge loss, defined as follows:

\begin{equation}
	\ell_{l h}=(1-Y \cdot \hat{Y})_{+}
\end{equation}

Where $ Y $  is the label, $ \hat{Y} $ is the predicted sample, $ (\cdot)_{+}=max(\cdot,0) $.

So mathematically, the hybrid segmentation loss can eventually be expressed as Eq. \ref{eq6}:

\begin{equation}	
	\begin{split}\ell_{s e g}(Y, \hat{Y})= -\frac{1}{N} \sum_{n=1}^{N}\Big(0.5 *\left(-\alpha\left(1-\hat{Y}_{n}\right)^{\beta} \log \left(\hat{Y}_{n}\right)\right)+ \\
		\frac{2 \cdot Y_{n} \cdot \hat{Y}_{n}+1}{Y_{n}+\hat{Y}_{n}+1}+0.5 *\left(Y_{n} \cdot \hat{Y}_{n}\right)_{+}\Big)
		\label{eq6}
	\end{split}
\end{equation}

Where $ Y_n \in Y $, $ \hat{Y}_n \in \hat{Y} $, respectively represents the $ n-th $ tiling GT picture and the tiling prediction probability picture, $ N $ represents the batch size.

\begin{figure*}[t]
	\centering
	\includegraphics[width=.95\linewidth]{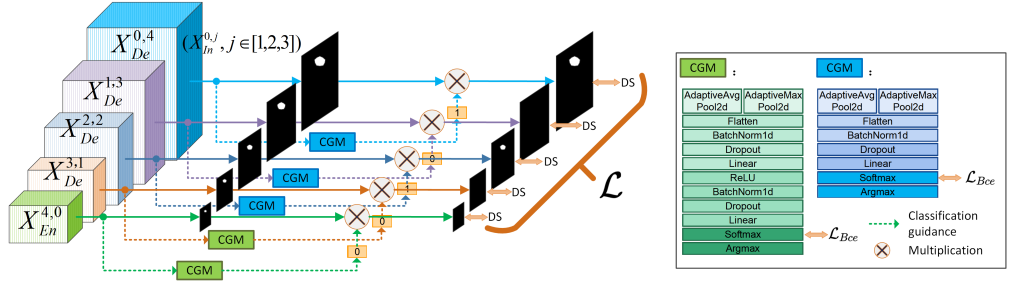}
	\caption{The principle of calculation for classification-guided module. The first layer nodes $ X_{In}^{0,j}, j \in [1,2,3] $ apply CGM in the same way as $X_{De}^{0,4}$, , so it is not drawn separately in the picture. The channel number of each node is displayed below the cuboids.}
	\label{fig3}
\end{figure*}

\subsection{Classification-guided module}
\label{sec4.3}

In the process of deep supervision calculation, some noisy information from the shallow layer background will lead to false positives, which will lead to over-segmentation, reducing the segmentation accuracy in the image segmentation task. Therefore, we refer to \cite{19} and propose a classification-guided module (CGM) applied in deep supervision. This module can classify the input image and judge whether it includes the organs or tissues to be segmented, avoiding the wrong segmentation caused by false positives to a certain extent, and improving the segmentation accuracy.

Different from the CGM structure proposed in UNet3+ \citep{19}, we propose two different CGM structures and apply them to the eight deep supervision branches of the model. As depicted in Figure \ref{fig3}, the green CGM structure is proposed to be applied in the deep supervision branches $X_{En}^{4,0}$ and $X_{De}^{3,1}$. Specifically, firstly, concatenates the global semantic features obtained through \textit{AdaptiveAvgPool2d} and \textit{AdaptiveMaxPool2d} operations. Then flattens the feature map by \textit{Flatten}, followed by the sequential operation including \textit{BatchNorm1d}, \textit{Dropout}, \textit{Linear}, \textit{ReLU}, and one more time of the sequential operation, will get a 2-channel feature map. Finally, using a \textit{Softmax} activation function generates a 2-dimensional tensor indicating the possibility of whether the input image has organs,  which realizes the classification that is optimized by the BCE loss function. Applying the \textit{Argmax} function yields the final output result composing of {0,1}, which is multiplied by the deep supervision branches, and finally gets the branch results after CGM operation. These results rectify some defects of over-segmentation caused by false positives. Considering that the channel number of $X_{De}^{2,2}$, $X_{De}^{1,3}$, $X_{De}^{0,4}$, and ${X_{In}^{0,j}},j \in[1,2,3]$ six branch feature maps are relatively small, we simplify the CGM structure shown with blue in Figure \ref{fig3}. It includes \textit{AdaptiveAvgPool2d}, \textit{AdaptiveMaxPool2d}, \textit{Flatten}, \textit{BatchNorm1d}, \textit{Dropout}, \textit{Linear}, \textit{Softmax},  \textit{Argmax} operations.

\subsection{Model pruning}
\label{sec4.4}

\begin{figure*}[t]
	\centering
	\includegraphics[width=1\linewidth]{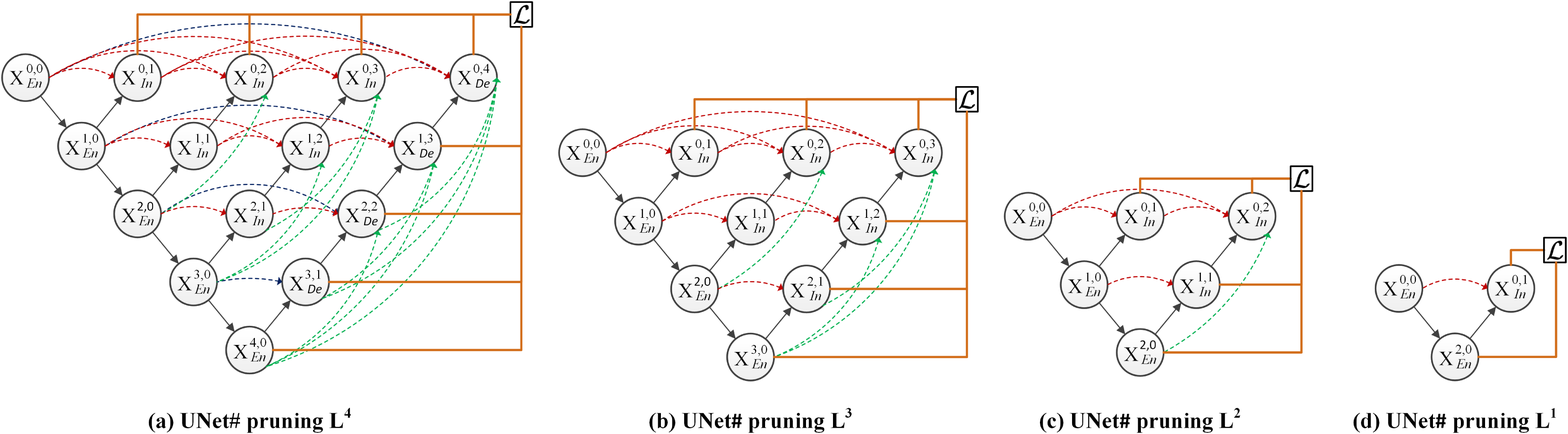}
	\caption{Illustration for the different levels of the proposed UNet\# pruning. Training the model with deep supervision prunes the model to $ L^4 $, $ L^3 $, $ L^2 $ and $ L^1 $ at the inference time. Model pruning $ L^4 $(a) indicates no pruning, whereas Model pruning $ L^1 $(d) represents the maximum pruning level, and its structure is the same as that of UNet2+ pruning of the same level. The calculation method of loss value $\mathcal{L} $ in each pruning level is similar, which takes the average loss value of all branches in the training time. See Figure \ref{fig2} and Sec. \ref{sec4.2} for specific diagrams and calculation process.}
	\label{fig4}
\end{figure*}

The introduction of deep supervision in our proposed network model makes the model have two operation modes (see Sec. \ref{sec4.2}), in which the fast mode can realize the model pruning function. As shown in Figure \ref{fig4}, the four pruning levels are signified as $ L^i,i\in[1,2,3,4] $. During the training phase of the model, the weight calculation includes both forward propagation and back-propagation, while in the test phase, the weight calculation only includes forward propagation calculation. Benefiting from the added deep supervision, the pruned branches of the model will have an impact on the remaining results due to back-propagation in the training stage; differently, they will not impact the output results in the inference stage. In other words, the $ L^4 $ level model participates training in the training phase, whose branches to be pruned contribute to the weight back-propagation. Meanwhile, in the inference time, according to the validation results, the model is pruned into the $ L^1 $, $ L^2 $, or $ L^3 $ level to reduce the network parameters, improving the network speed (Pruning $ L^4 $, a full model means no pruning in the inference).

\section{Experiments and results}
\label{sec5}

\subsection{Datasets}

To evaluate the performance of the proposed model more comprehensively, we consider using datasets with different modalities and organs, specifically including Dsb2018 (Nuclei segmentation) , BraTs19 (Brain tumor segmentation), Lits17 (Liver segmentation) , LIDC-IDRI (Lung 2D segmentation), Luna16 (Lung 3D segmentation), wherein the first four datasets are used for the training and validation of the 2D network model and Luna16 dataset is used for the 3D network model. All these datasets will be separately divided into training set, validation set, and test set in the experiment.

\subsection{Evaluation metrics}
\label{sec5.1}
In our experiments, Dice-Coefficient (DC) and Intersection over Union (IoU) are used as the evaluation metrics of model performance. DC is a set similarity measurement function that calculates the pixel-wise similarity between the predicted image and the GT. IoU is a standard for measuring the accuracy of detecting corresponding objects in a specific dataset, representing the pixel-wise overlap rate or degree of the prediction and GT, namely, their intersection to union ratio.

\subsection{Implementation details}
\label{sec5.2}
We preprocess the datasets with \textit{RandomRotate90}, \textit{Flip}, a random selection (HSV color space transformation, brightness contrast change), and \textit{Normalization} to train the models. \textit{Adam} optimization function with weight\_decay 1e-4 and small batch sizes of 128 samples to calculate gradient updates in training. The network's weights are initialized by \cite{74}, trained for 100 passes using the learning rate (initialized to 1e-3) adjusted by \textit{CosineAnnealing} \citep{75}. We code the experiments on the PyTorch open-source framework and log the experiment data by the open-source tool Wandb. Both training and validation of all models are parallel conducted on two NVIDIA Tesla A100 GPU with 80GB memory in a 24-core, 48 threads server equipped with an Intel Xeon Silver 4214R 2.4GHz CPU(128GB RAM).

\begin{table*}[t]
	\footnotesize
	\caption{Evaluation metric IoU (\%) of semantic segmentation results for UNet, WUNet, UNet2+, UNet3+, and the proposed UNet\# on nuclei, brain tumor, liver, 2D lung node, and 3D lung node. The model backbone is set as VGG11. DS is the abbreviation of deep supervision, $ \times $ indicates without DS, and $ \checkmark $ means with DS. Among all the evaluation results, the best result is shown in \textbf{bold}.}
	\label{table2}
	\centering	
	\setlength{\tabcolsep}{0.8mm}{
		\begin{tabular}{lcc|cccc|lcc|c}
			\toprule
			\multirow{2}{*}{model} & \multirow{2}{*}{DS} & \multirow{2}{*}{Params} & \multicolumn{4}{c|}{2D segmentation}                                                                  & \multirow{2}{*}{model} & \multirow{2}{*}{DS} & \multirow{2}{*}{Params} & 3D segmentation \\ \cmidrule{4-7} \cmidrule{11-11} 
			&                     &                         & \multicolumn{1}{c}{Dsb2018} & \multicolumn{1}{c}{BraTs19} & \multicolumn{1}{c}{Lits17} & LIDC-IDRI &                        &                     &                         & Luna16          \\ \midrule
			UNet                   & $ \times $                & 7.85M                  & \multicolumn{1}{c}{90.15}  & \multicolumn{1}{c}{89.96}   & \multicolumn{1}{c}{91.43}  & 72.59     & V-Net     & $ \times $                 & 26.39M                 & 71.21          \\ 
			WUNet                  & $ \times $                   & 9.39M                   & \multicolumn{1}{c}{90.27}   & \multicolumn{1}{c}{90.06}   & \multicolumn{1}{c}{91.52}  & 72.93     & WV-Net                 & $ \times $                   & 33.40M                  & 73.52           \\ 
			UNet2+                  & $ \times $                   & 9.16M                   & \multicolumn{1}{c}{92.02}   & \multicolumn{1}{c}{90.03}   & \multicolumn{1}{c}{91.59}  & 72.73     & V-Net2+                & $ \times $                   & 27.37M                  & 76.35           \\ 
			UNet2+                 &  $ \checkmark $                 & 9.16M                   & \multicolumn{1}{c}{91.98}   & \multicolumn{1}{c}{91.01}   & \multicolumn{1}{c}{91.64}  & 73.43     & V-Net2+                & $ \checkmark $                   & 27.37M                  & 77.62           \\ 
			UNet3+                 & $ \times $                   & 9.79M                   & \multicolumn{1}{c}{91.96}   & \multicolumn{1}{c}{90.67}   & \multicolumn{1}{c}{91.21}  & 73.22     & V-Net3+                & $ \times $                   & 27.84M                  & 76.78           \\ 
			UNet3+               & $ \checkmark $                   & 9.79M                   & \multicolumn{1}{c}{92.15}   & \multicolumn{1}{c}{91.69}   & \multicolumn{1}{c}{92.14}  & 73.48     & V-Net3+                & $ \checkmark $                  & 27.84M                  & 78.23           \\ 
			UNet\#                 & $ \times $                   & 9.71M                   & \multicolumn{1}{c}{92.53}   & \multicolumn{1}{c}{91.67}   & \multicolumn{1}{c}{94.79}  & 73.68     & V-Net\#                & $ \times $                   & 28.37M                  & 77.69           \\ 
			UNet\#                 & $ \checkmark $                   & 9.71M                   & \multicolumn{1}{c}{\textbf{92.67}}   & \multicolumn{1}{c}{\textbf{92.38}}   & \multicolumn{1}{c}{\textbf{95.36}}  & \textbf{74.01}     & V-Net\#                & $ \checkmark $                   & 28.37M                  & \textbf{79.45}           \\ \bottomrule
		\end{tabular}
	}
	
\end{table*}

\begin{figure*}[t]
	\centering
	\includegraphics[width=.95\linewidth]{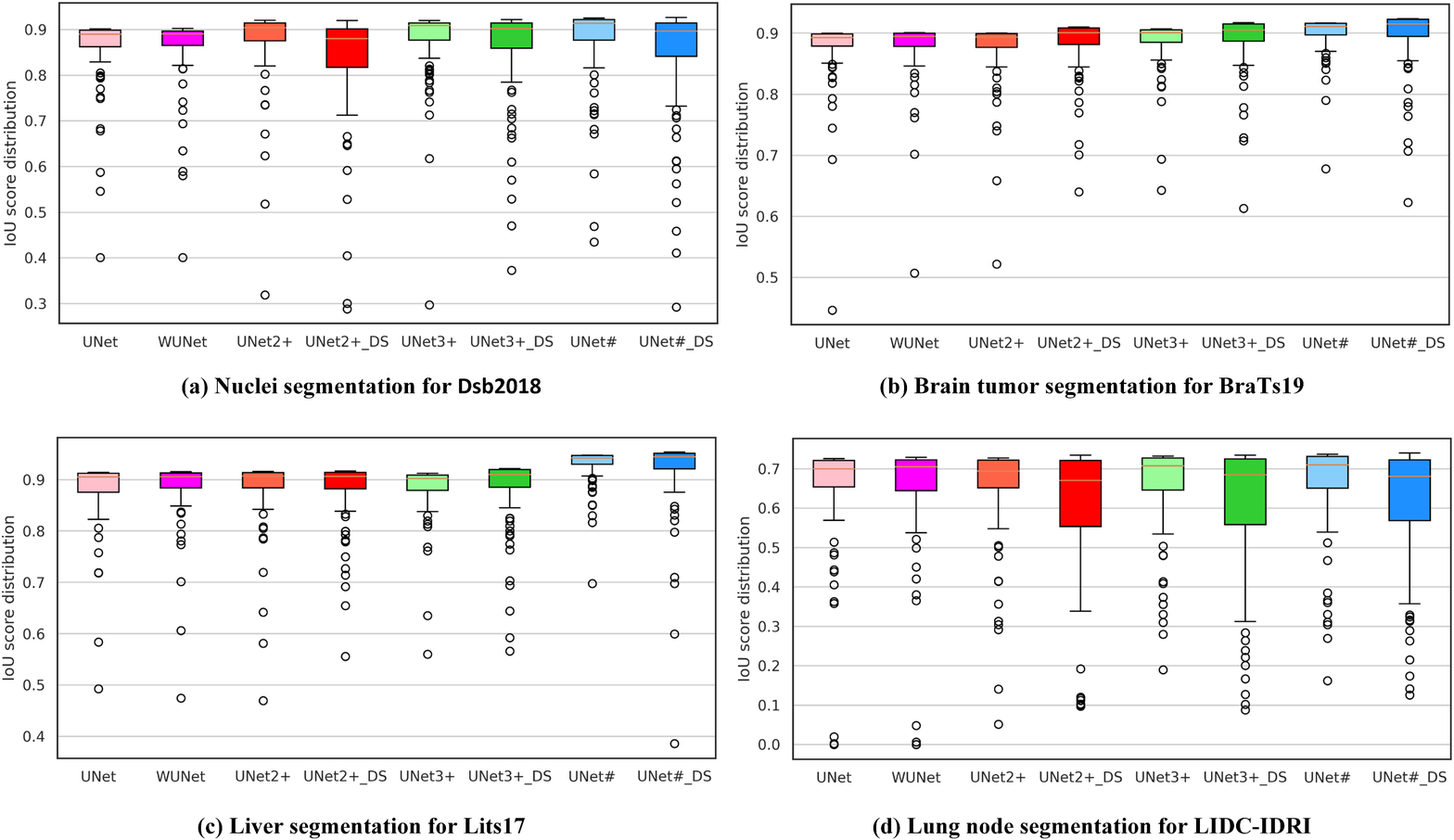}
	\caption{The boxplots of validation results for UNet, WUNet, UNet2+, UNet3+, and the proposed UNet\# on 2D datasets Dsb2018(a), BraTs19(b), Lits17(c), and LIDC-IDRI(d). The IoU boxplots show the scores of the quartile ranges, whiskers and dots indicate outliers on the validation datasets. UNet2+\_DS, UNet3+\_DS, UNet\#\_DS indicates the corresponding model with deep supervision.}
	\label{fig5}
\end{figure*}

\subsection{Results}
\label{sec5.3}
The experimental results of all models are obtained by running in a fair and unbiased environment, which are visually presented in pictures and tables. 

\subsubsection{Comparison on semantic segmentation}
\label{sec5.3.1}
This section compares the semantic segmentation performance of the proposed UNet\# with UNet, wide-UNet (WUNet), UNet2+, and UNet3+ on the datasets. We analyze the experiment results from both qualitative and quantitative aspects.

\textbf{Quantitative analysis}: Table \ref{table2} detailly summarizes the parameters of each model with VGG11 as the backbone and their IoU results on the five datasets. The number of parameters for WUNet is 9.39M, reaching 1.54M more than UNet. Benefitting from more parameters, WUNet gets more significant performance on each dataset than UNet. However, unexpectedly, this superiority does not work compared with UNet2+ and UNet3+. This situation shows that simply raising channels for the performance improvement of the model is not better than optimizing the model structure. The proposed UNet\# has 9.71M parameters, 1.68M more than UNet, and 0.32M more than WUNet. Nevertheless, unlike WUNet, UNet\# is optimized and improved at the model structure by redesigning the skip connections, resulting in heavier parameters but better segmentation metrics. In the nuclei dataset (Dsb2018), UNet\# respectively achieves an IoU gain of 0.51 and 0.57 points over UNet2+ and UNet3+. Similarly, it performs a satisfied IoU gain over UNet2+ and UNet3+ across all the other four tasks of brain tumor ($ \uparrow $1.64, $ \uparrow $1.0), liver ($ \uparrow $3.2, $ \uparrow $3.58), 2D lung nodule ($ \uparrow $0.95, $ \uparrow $0.46) segmentation. Adding deep supervision to UNet\# further performs a distinct effect on all datasets.

Figure \ref{fig5} shows the validation IoU score of each model on the four 2D datasets during 100 pass training. It can be seen from the figure that the proposed model and the model with deep supervision have better segmentation performance for organs and lesions than other models, wherein Figure \ref{fig5} (c) has particularly obvious advantages in liver segmentation.

Furthermore, we compare the IoU results of the V-Net \citep{76} model applied ++ \citep{18}, +++ \citep{19}, and \# (our method) schemes on the 3D lung dataset. The feature list of V-Net is [16,32,64,128,256], which is slightly expands to [18,36,72,144,288], getting wide V-Net (WV-Net) whose parameters are 33.4M. The heavier parameter cannot bring the best result, indicating again that simply adding features may inhibit the learning ability of the model. V-Net\# achieving the best IoU of 77.69\% (with DS 79.45\%) shows that integrating dense skip connections and full-scale skip connections can promote the model performance at a small increased cost.

\begin{figure*}[t]
	\centering
	\includegraphics[width=.95\linewidth]{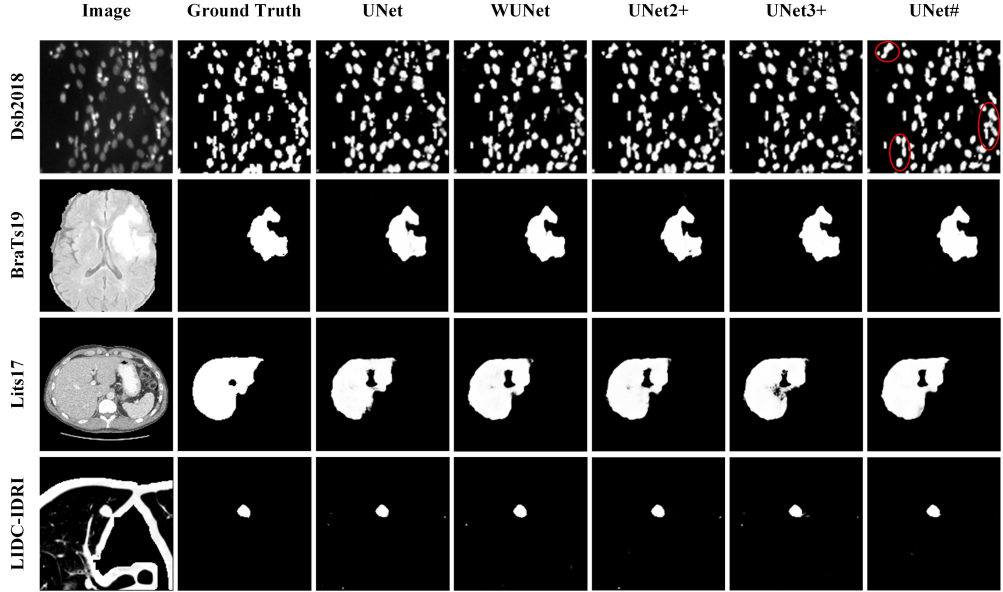}
	\caption{Qualitative comparison of 2d segmentation results between UNet, WUNet, UNet2+, UNet3+, and UNet\# on Dsb2018, BraTs19, Lits17, and LIDC-IDRI datasets. The last image of segmentation result from the first line, in which the red circle is added in anaphase, clearly indicates the place where the segmentation result is better.}
	\label{fig6}
\end{figure*}

\textbf{Qualitative analysis}: Figure \ref{fig6} exhibits the segmentation results of the models in each dataset. The first column is the original image for testing, followed by ground truth (GT), and the third to seventh columns are the test results for UNet, WUNet, UNet2+, UNet3+, and the proposed UNet\#. As shown by the red circle in the last column of the first row, testing in the Dsb2018 dataset, the segmentation results obtained by UNet\# are closer to the GT than other models. Similarly, it is noticeable that UNet\ achieves consistent high-performance results in the localization and segmentation of organs and lesions on other datasets.

\subsubsection{Comparison on instance segmentation}
\label{sec5.3.2}

In addition to segmenting the boundary of each instance on the pixel, instance segmentation also needs to frame different object instances with the target detection method, so it is more challenging than semantic segmentation. Since the images in the Dsb2018 dataset contain multiple nuclei of cells, suitable for the challenge of instance segmentation, this section details the instance segmentation results of different methods on the Dsb2018.

In this experiment, to illustrate that the proposed method is a backbone-agnostic extension to UNet, we perform the models with VGG16 and ResNet101 as the backbone. Mask R-CNN \citep{77} is a classic model in instance segmentation, including feature pyramid network (FPN) structural blocks, which can generate target recommendations on multiple scales. Taking it as the benchmark network [6], we redesign the plain skip connections of FPN with ++ \citep{18}, +++ \citep{19}, and \# (our method) schemes, generating Mask R-CNN2+, Mask R-CNN3+, Mask R-CNN\# for the instance segmentation experiment in this section.

The second and third rows of Table \ref{table3} compare semantic segmentation experimental results of UNet and UNet\# on the Dsb2018 dataset. Based on the backbone of VGG16, the proposed UNet\# achieves 92.82\% IoU and 89.60\% Dice, which are respectively enhanced by 2.7 points (vs. 90.12\%) and 2.66 points (vs. 86.94\%) compared with UNet; Furthermore, the experimental results based on resnet101 backbone demonstrate that our method achieves consistent performance. The fourth to seventh rows of Table \ref{table3} compares the experimental results of the instance segmentation on Dsb2018. It can be seen that the Mask R-CNN of our scheme (Mask R-CNN\#) has achieved the best segmentation results. In conclusion, as expected, our method is a backbone-agnostic optimization scheme, outperforming UNet, ++, +++ method in semantic segmentation and instance segmentation.

\begin{table*}[t]
	\caption{Comparison of IoU (\%) and Dice (\%) for the semantic and instance segmentation tasks. The upper part of the table is the validation results of nuclear semantic segmentation for UNet and UNet\# on Dsb2018 dataset; The lower part is the validation results of nuclear instance segmentation for Mask R-CNN \citep{77}, Mask R-CNN2+, Mask R-CNN3+ and Mask R-CNN\# on Dsb2018. $ \dag  $ Mask R-CNN with ++ \citep{18}, +++ \citep{19}, and \# (our method) schemes design in its feature pyramid. The best results are shown as \textbf{bold}}
	\label{table3}
	\centering
	\setlength{\tabcolsep}{3.5mm}{
		\begin{tabular}{l|ccc|ccc}
			\toprule
			model               & Backbone & IoU   & Dice  & Backbone  & IoU   & Dice  \\ \midrule
			UNet                & VGG16    & 90.12 & 86.94 & ResNet101 & 91.52 & 89.43 \\ 
			UNet\#              & VGG16    & \textbf{92.82} & \textbf{89.60} & ResNet101 & \textbf{93.56} & \textbf{90.78} \\ \midrule
			Mask R-CNN  & VGG16    & 92.34 & 87.68 & ResNet101 & 93.45 & 89.34 \\ 
			Mask R-CNN2+$ \dag  $      & VGG16    & 93.47 & 88.14 & ResNet101 & 94.96 & 90.16 \\ 
			Mask R-CNN3+ $ \dag  $       & VGG16    & 94.28 & 90.51 & ResNet101 & 95.26 & 91.38 \\ 
			Mask R-CNN\#  $ \dag  $      & VGG16    &\textbf{ 94.89} & \textbf{91.37} & ResNet101 & \textbf{95.87} & \textbf{92.25} \\ \bottomrule
	\end{tabular}}
\end{table*}

\subsubsection{Comparison on model pruning}
\label{sec5.3.3}
The proposed model designs two types of deep supervision for branch pruning. One is adding a loss function to the output of four decoder branches for deep supervision to learn the hierarchical representation from the full-scale aggregation feature map. The other is performing deep supervision on the four branches at the first level layer, whose scale is the same as the input image size, to realize the branch pruning. 

Actually, there are two ways to train the model in implementing model pruning. The first method is called embedded training, specifically comprehended as training a high pruning-level UNet\# and then pruning the model to get a low pruning-level model. For example, the model calculates $X_ {In}^{0,1}$, $X_ {In}^{0,2}$, $X_ {in} ^ {0,3} $and $X_ {de} ^ {0,4} $ node loss, supervising training the full UNet\# (equivalent to pruning-level $L^4$), and then pruning at level \textit{i} to get UNet\# $L^i, i \in [1,2,3]$; Or for another example, the model calculates $X_ {In}^{0,1}$, $X_ {In}^{0,2}$, $X_ {in} ^ {0,3} $ node loss supervising training UNet\# $ L^3 $, and then pruning model to obtain UNet\# $L^i, i \in [1,2]$. Embedded training can be regarded as integrating sub-networks training at different depths. The second method is isolated training, specifically understood as training the pruning $ L^i $ model independently and using it for inference. For example, in the training phase, UNet\# pruning $ L^3 $ is trained without interactions with the deeper encoder and decoder nodes from $ L^4 $ and is used for inference without further pruning. Comparison of this two training methods are detailed in \ref{Appendix B}. The experimental results in this section are performed with the first training method.

\begin{figure*}[t]
	\centering
	\setlength{\belowcaptionskip}{-0.5cm}
	\includegraphics[width=.95\linewidth]{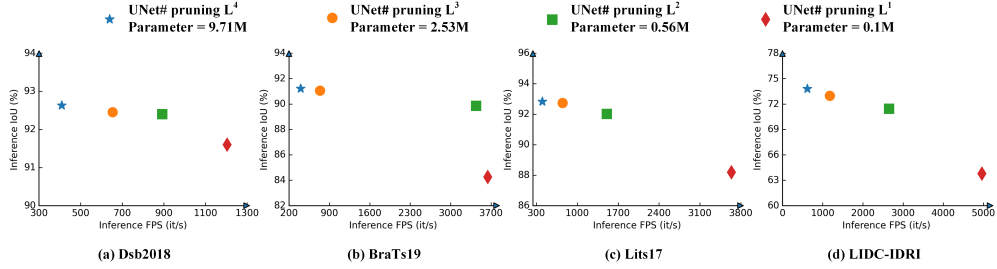}
	\caption{Illustration of different model pruning level. Model complexity, inference speed represented by Frame Per Second (FPS), and inference IoU (\%) of UNet\# after pruning on Dsb2018(a), BraTs19(b), Lits17(c), LIDC-IDRI(d) datsets. The inference speed of the experiments is carried out in the same hardware enviroment detailed in Sec. \ref{sec5.2}.}
	\label{fig7}
\end{figure*}

Figure \ref{fig7} shows the test results of four different levels of pruning models on four various datasets. As seen, branch pruning generates a wide margin in the number of model parameters at different levels; specifically, $ L^1 $ is 0.1M, 5 times less than $ L^2 $ (vs. 0.56m), 25 times less than $ L^3 $ (vs. 2.53m), and 97 times less than $ L^4 $ (vs. 9.71m). $ L^1 $ parameters are the least but with the worst results, demonstrating that a too shallow model can not achieve satisfactory results. Nevertheless, more importantly, considering the expensive calculation cost and memory intensive of the existing deep convolution neural networks, transplanting lighter pruned models to mobile devices can promote the daily application and comprehensive development of computer-aided diagnosis (CAD).

\subsubsection{ Comparison with the State-of-the-Art}
\label{sec5.3.4}

We make a quantitative comparison between the proposed method and other state-of-the-art methods including PSPNet \citep{5}, DeepLabV3+ \citep{38}, ENet\citep{26}, nnU-Net \citep{78}, AttnUNet \citep{79}, DUNet \citep{80}, Ra-UNet \citep{81}, and TransUNet \citep{82} on BraTs19 and Lits17 datasets according to the fair comparison of super parameters and training environmental configuration.

\begin{table}[h]
	\centering
	\caption{Comparison of UNet\# and other 8 state-of-the-art models on BraTs19 and Lits17 datasets. UNet\# with calssification-guided module (CGM) achieves the best results highlighted in \textbf{bold}.}
	\label{table4}
		\begin{tabular}{l|cc|cc}
			\toprule
			\multirow{2}[2]{*}{model} & \multicolumn{2}{c|}{BraTs19} & \multicolumn{2}{c}{Lits17} \\ \cmidrule{2-5} 
			\multicolumn{1}{c|}{} & \multicolumn{1}{c}{IoU} & \multicolumn{1}{c|}{Dice} & \multicolumn{1}{c}{IoU} & \multicolumn{1}{c}{Dice} \\
			\midrule
			PSPNet \citep{5} & 83.85 & 91.01 & 87.75 & 93.23 \\
			DeepLabV3+ \citep{38}& 87.1  & 92.9  & 91.07 & 95.17 \\
			Enet  \citep{26}& 81.62 & 89.9  & 83.36 & 90.9 \\
			nnU-Net \citep{78}& 88.52 & 91.68 & 91.31 & 95.3 \\
			AttnUNet \citep{79}& 91.32 & 95.32 & 94.61 & 97.15 \\
			DUNet \citep{80}& 90.86 & 95.19 & 94.47 & 97.21 \\
			Ra-Unet \citep{81}& 84.28 & 89.12 & 91.94 & 96.1 \\
			TransUNet \citep{82}& 88.85 & 94.04 & 92.59 & 96.16 \\
			UNet\# & 91.68 & 95.6  & 94.79 & 97.29 \\
			UNet\#(CGM) & \textbf{92.08} & \textbf{96.14} & \textbf{95.31} & \textbf{98.24} \\
			\bottomrule
	\end{tabular}
\end{table}%

As seen from Table \ref{table4}, Our method earns better IoU (91.68\%, 94.79\%) and Dice (95.60\%, 97.29\%) metrics than other methods in the two datasets. Embedding the CGM module can achieve 0.4, 0.52 gains on IoU, and 0.54, 0.95 gains on Dice. These demonstrate that the mean of improving UNet in this paper is feasible, and the improvement of skip-connections is significant. Meanwhile, it also confirms that compared with Separable pyramid pooling methods \citep{5,38}, NLP Transformers combination UNet \citep{82}, Attention UNet \citep{79,81}, and deformable convolution combining with UNet \citep{80}, the proposed UNet\# is sophisticated in medical image segmentation.
\section{Conclusion}
\label{sec7}
Inspired by the successful improvement of skip connections in UNet2+ and UNet3+, this paper proposes a new model structure named UNet-sharp (UNet\#). The proposed model aggregates the features of different semantic scales on the encoder and decoder subnet through nested and dense skip connections. Furthermore, it utilizes full-scale connections to maximize the fusion of high-level semantics and low-level details in multi-scale feature maps to realize accurate organ or lesion location perception and segmentation. Introducing deep supervision into the eight branches can make the model have a pruning function and learn the hierarchical representation from the full-scale aggregation feature map. Additionally, two kinds of CGM are designed for the model to avoid over-segmentation. Experiments from semantic and instance segmentation on different organ and pattern datasets demonstrate that the proposed model outperforms all the state-of-the-art methods.


\section*{Acknowledgments}
This work was supported in part by Wenzhou Association For Science and Technology under [grant no. kjfw39], in part by the Major Project of Wenzhou Natural Science Foundation under [grant no. ZY2019020], in part by the Project of Wenzhou Key Laboratory Foundation under [grant no. 2021HZSY0071], in part by the Department of Education of Zhejiang Province under [grant no. Y202146494], in part by the Soft Science Key Research Project of Zhejiang Province under [grant no. 2022C25033] and in part by the Key Project of Zhejiang Provincial Natural Science Foundation under [grant no. LD21F020001].

\newpage

\bibliographystyle{plainnat}
\bibliography{main}

\newpage




\end{document}